\documentclass[aps, prd, amsmath, floats, floatfix, twocolumn, nofootinbib, superscriptaddress, %showpacs
]{revtex4-1}
\usepackage{graphicx}
\usepackage{latexsym}
\usepackage{color}

\newcommand{\be}{\begin{eqnarray}}
\newcommand{\ee}{\end{eqnarray}}
\newcommand{\rar}{\rightarrow}

\begin{document}

\title{Lee-Wick Black Holes}

\author{Cosimo Bambi}
\email{bambi@fudan.edu.cn}
\affiliation{Center for Field Theory and Particle Physics and Department of Physics, Fudan University, 200433 Shanghai, China}
\affiliation{Theoretical Astrophysics, Eberhard-Karls Universit\"at T\"ubingen, 72076 T\"ubingen, Germany}

\author{Leonardo Modesto}
\email{lmodesto@sustc.edu.cn}
\affiliation{Department of Physics, Southern University of Science and Technology, Shenzhen 518055, China}

\author{Yixu Wang}
\email{yw417@cam.ac.uk}
\affiliation{Department of Applied Mathematics and Theoretical Physics, University of Cambridge, Cambridge CB3 0WA, United Kingdom}

\date{\today}

\begin{abstract}
We derive and study an approximate static vacuum solution generated by a point-like source in a higher derivative gravitational theory with a pair of complex conjugate ghosts. The gravitational theory is local and characterized by a high derivative operator compatible with Lee-Wick unitarity. In particular, the tree-level two-point function only shows a pair of complex conjugate poles besides the massless spin two graviton. We show that singularity-free black holes exist when the mass of the source $M$ exceeds a critical value $M_{\rm crit}$. For $M > M_{\rm crit}$ the spacetime structure is characterized by an outer event horizon and an inner Cauchy horizon, while for $M = M_{\rm crit}$ we have an extremal black hole with vanishing Hawking temperature. The evaporation process leads to a remnant that approaches the zero-temperature extremal black hole state in an infinite amount of time.
\end{abstract}

\maketitle

%%%%%%%%%%%%%%%%%%%%%%%%%%%%%%%

\section{Introduction \label{s-1}}

The quantization problems of the Einstein-Hilbert action are well known. In the past 40 years, many authors have tried to quantize gravity by introducing modifications to Einstein's gravity. The first higher derivative theory of gravity dates back to quadratic gravity, which was proposed by Stelle in 1977~\cite{stelle}. Stelle's theory is renormalizable and asymptotically free, but it is not unitary, having a massive ghost state in the spectrum.

A class of self-consistent quantum theories is represented by weakly non-local modifications of Einstein's gravity. These theories were first discussed by Krasnikov and Kuz'min~\cite{krasnikov,kuzmin}, following previous work by Efimov~\cite{efimov}. A significant contribution to this line of research was later provided by Tomboulis, who proposed a whole class of weakly non-local super-renormalizable gauge and gravitational theories~\cite{tom1,tom2,tom3}. Recently, there have been a renewed interest in this class of gravity theories, which have been extensively studied to better understand their quantum properties~\cite{lm1,lm2}. In particular, a simple extension of~\cite{lm3} turns out to be completely finite at quantum level. Preliminary studies of black holes and gravitational collapse in these theories are reported in~\cite{cb1,cb2,Frolov3,Frolov4,Frolov5,Frolov6,Frolov7,Frolov8}.

The weakly nonlocal theories are a quasi-polynomial extension of the higher derivative theories introduced and studied by Asorey, Lopez, Shapiro in~\cite{shapiro1}. Recently, Shapiro has pointed out that the quantum effective action of weakly nonlocal theories has likely an infinite number of complex conjugate poles~\cite{shapiro2}. Therefore, we decided to come back to the local higher derivatives theories (beyond Stelle's theory) with the special property of admitting the graviton field and only complex conjugate poles (no real poles) in the classical spectrum~\cite{shapiro2,shapiro3}. These theories are unitary in agreement with the Lee-Wick prescription~\cite{leewick1,leewick2,CLOP}.

In this paper, we focus on the minimal theory that fulfills the properties listed above. The action is~\cite{shapiro3, lm5,giacchini,deBrito:2016fqe,shapiro5}:
\be 
S =  2 \kappa^{-2} \int \! d^4 x \sqrt{|g|} \, 
\Big[ R  + \alpha_g^2 \, G_{\mu \nu} \Box R^{\mu \nu} \Big] \,  ,
 \label{4Daction2}
 \ee 
where $\kappa^2 = 16 \pi G_{\rm N}$, $\alpha_g = 1/\Lambda^2$, and $\Lambda$ is the UV scale of the theory. $\Lambda$ is not necessarily equal to the Planck mass, but it may be expected of the same order as the Planck mass.

Looking at the exact equations of motion (EOM), we can immediately infer that all Ricci-flat spacetimes are exact solutions of the theory in vacuum \cite{yaudong}. However, when a point-like source is introduced on the right side of the EOM, the Schwarzschild, the Kerr, and other Ricci-flat spacetimes are no more exact solutions. Indeed, the Newtonian potential turns out to be regular and constant near $r=0$ in any general higher derivative theory~\cite{tiberio,shapiro4,shapiro5} (in nonlocal gravity we have a similar regular behaviour \cite{biswas}). We thus expect a similar regular behaviour also for the exact black hole solutions, if any, when the EOM are solved in a non-empty spacetime. On the footprint of these results, in this paper we only consider approximate EOM that we can somehow ``solve exactly", namely 
\be\label{eq-e}
\left(1+ \frac{\Box^2}{\Lambda^4}  \right)G_{\mu\nu} + O ( R_{\mu\nu}^2 )
= 8 \pi G_{\rm N} \, T_{\mu\nu} \, .
\label{AEOM}
\ee

\section{Black hole solutions \label{s-2}}

Let us consider a static point-like source of mass $M$. The only non-vanishing component of its energy-momentum tensor is:
\be\label{eq-t}
T^0_0 = - M \delta ({\vec x}) \, ,
\ee
where $\delta(\vec{x})$ is the Dirac delta function and ${\vec x} = (x,y,z)$ are the Cartesian coordinates of the 3-space. Eq.~(\ref{AEOM}) can be interpreted as the standard Einstein equations with an effective matter source on the left hand side. The effective energy-momentum tensor is 
\be
\tilde{T}_{\mu\nu} \approx 
\left( 1 + \frac{\Box^2}{\Lambda^4} \right)^{-1} T_{\mu\nu} \, . 
\ee
The approximate EOM we are going to solve (leaving aside operators $O(R^2)$) read
\be
\label{eq-e2}
%\hspace{-0.5cm}
G_{\mu\nu} = %+ O ( R_{\mu\nu}^2 ) &=& 
8 \pi G_{\rm N}  \tilde{T}_{\mu\nu} %\nonumber\\
= 8 \pi G_{\rm N}  \left( 1 + \frac{\Box^2}{\Lambda^4}   \right)^{-1} T_{\mu\nu}  . 
\ee
We point out that this is a drastic approximation of the exact EOM coming from the theory~(\ref{4Daction2}), but the outcome will turn out to be consistent with the results obtained in the Newtonian approximation~\cite{tiberio,shapiro4,shapiro5}. Moreover, as we have already pointed out, the Ricci-flat solutions are mathematically inconsistent in presence of a point-like source.

With the choice~(\ref{eq-t}), the effective energy-momentum tensor can be written as
\be\label{eq-tmunud}
\tilde{T}^\mu_\nu = {\rm diag} \left ( - \tilde{\rho} , \tilde{P}_r , 
\tilde{P}_\theta , \tilde{P}_\theta \right) \, ,
\ee
where $\tilde{\rho}$ is the effective energy density, $\tilde{P}_r$ is the effective radial pressure, and $\tilde{P}_\theta$ is the effective tangential pressure. The effective energy density is
\be\label{eq-rho}
&&\hspace{-0.6cm}
 \tilde{\rho}(r) = \left( 1 + \frac{\Box^2}{\Lambda^4} \right)^{-1} M \delta ({\vec x}) = M \int \frac{dk^3}{(2\pi)^3} \frac{e^{i \vec k \vec x}}{1+ (k/\Lambda)^4} \nonumber \\
&& \hspace{0.15cm} = \frac{M \Lambda^2}{4 \pi r} e^{-\frac{r\Lambda}{\sqrt 2}} \sin\frac{r\Lambda}{\sqrt 2} \, .
\ee

Let us assume that the static and spherically symmetric solution of Eq.~(\ref{eq-e}) has the usual Schwarzschild-like form
\be\label{eq-ds}
ds^2 = -F(r)dt^2 + \frac{dr^2}{F(r)} + r^2 d\Omega^2 \, , 
\ee
where
\be
F(r) = 1- \frac{2G_{\rm N} m(r)}{r} \, .
\ee
$m(r)$ is some effective mass and is a function of the radial coordinate $r$ only because of the spherical symmetry.

With the energy-momentum tensor in Eq.~(\ref{eq-tmunud}) and the metric ansatz in~(\ref{eq-ds}), the Einstein equations (\ref{eq-e2}) turn into 
\be
\label{eq-ee-1}
\frac{dm}{dr} &=& 4 \pi r^2 \tilde{\rho} \, , \\
\label{eq-ee-2}
\frac{1}{F} \frac{dF}{dr} &=& \frac{2 G_{\rm N} 
\left( m + 4 \pi r^3 \tilde{P}_r \right)}{r \left( r - 2 G_{\rm N} m \right) } \, , \\
\label{eq-ee-3}
\frac{d \tilde{P}_r}{dr} &=& - \frac{1}{2F} \frac{dF}{dr} \left( \tilde{\rho} + \tilde{P}_r\right) 
+ \frac{2}{r} \left( \tilde{P}_\theta - \tilde{P}_r \right) \, .
\ee
From Eq.~(\ref{eq-ee-1}), we find the function $m$
\be\label{eq-dmdr}
m (r) = 4 \pi \int_0^r dx \, x^2 \tilde{\rho}(x) \, .
\ee
Eq.~(\ref{eq-ee-2}) is solved by $\tilde{P}_r = - \tilde{\rho}$, while From Eq.~(\ref{eq-ee-3}) we derive $\tilde{P}_\theta$.

\begin{figure*}[t]
\begin{center}
\vspace{0.8cm}
\boxed{
\includegraphics[width=7.5cm]{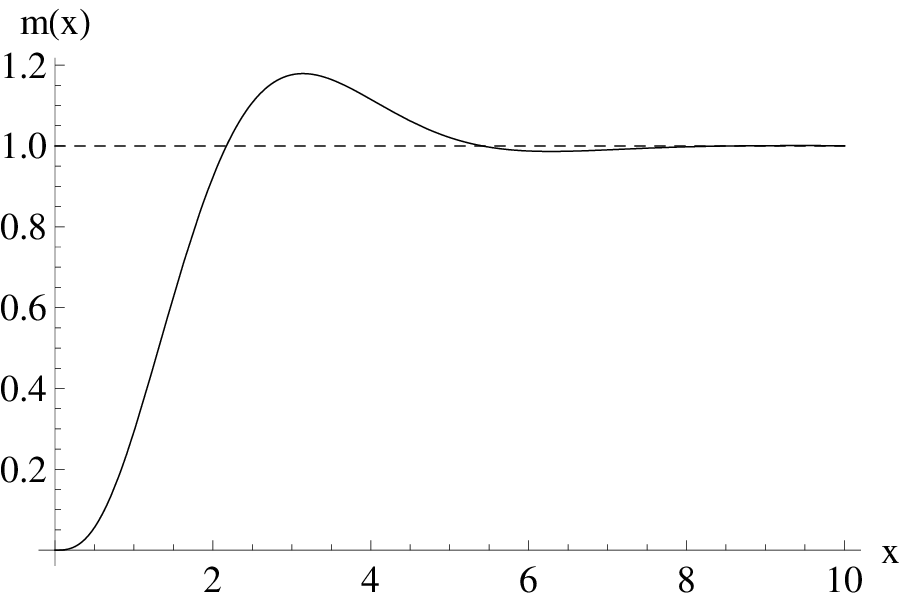}}
\hspace{1.2cm}
\boxed{\includegraphics[width=8.1cm]{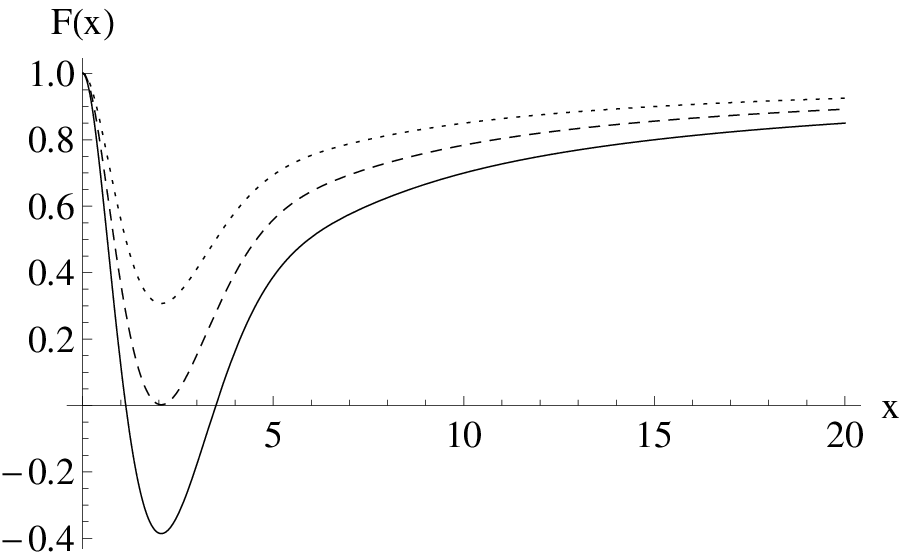}}
\end{center}
\caption{Left panel: effective mass $m(x)$ for $M=1$. Right panel: $F(x)$ for $M = 1.5$ (dotted line, no horizon), $M=2.165$ (dashed line, one horizon), and $M=3$ (solid line, two horizons). See the text for more details. \label{fig1}}
\end{figure*}

If we plug the effective energy density~(\ref{eq-rho}) into Eq.~(\ref{eq-dmdr}) and we integrate over the radial coordinate $r$, we find
\be
m(x) = M f(x) \, ,
\ee
where $x = \Lambda r / \sqrt{2}$ is a dimensionless coordinate and $f(x)$ is the dimensionless effective mass
\be
f(x) = 1 - e^{-x} \left[ (1+x) \cos x + x \sin x \right] \, .
\ee
The left panel in Fig.~\ref{fig1} shows the profile of $m(x)$ for $M=1$. For $x \gg 1$, we recover the limit of general relativity with $m = M$ and the metric reduces to the Schwarzschild solution. At small radii, there are deviations from the classical picture. The characteristic length scale is $1/\Lambda$, which is the UV cut-off of the theory and is presumably extremely small, like the Planck length even if it is not necessarily the Planck length. $m(x)$ has a bump at $x \approx 3$ because the effective energy density is negative between $x \approx 3$ and $x \approx 5$. In other words, if we interpret this model as Einstein's gravity coupled to an effective energy-momentum tensor rather than as a non-local modification of Einstein's gravity, the effective energy-momentum tensor violates some energy conditions.

If we expand the effective mass around the centre $r = 0$, we have 
\be
m = \frac{\Lambda^3 r^3}{3 \sqrt{2}} \, M + ... \, .
\ee
We thus find a de~Sitter core
\be
\hspace{-0.8cm}
ds^2 \approx -\left(1- \frac{\Lambda_{\rm cc}}{3} r^2 \right) dt^2
+ \frac{dr^2}{1- \frac{\Lambda_{\rm cc}}{3} r^2}
+ r^2 d\Omega^2 \, , 
\ee
where 
$\Lambda_{\rm cc} = \sqrt{2} G_{\rm N} \Lambda^3 M$ is the effective cosmological constant of the de~Sitter spacetime. Since the spacetime is de~Sitter  at the center, it is regular. Scalar curvatures like the Kretschmann do not diverge and the spacetime is geodetically complete. If $\Lambda$ is of the order of the Planck mass $M_{\rm Pl}$, $\Lambda_{\rm cc} \sim M_{\rm Pl} M$. For $M \gg M_{\rm Pl}$ (e.g. astrophysical black holes), the de~Sitter core is characterized by a huge value of the effective cosmological constant. While this does not directly represent a problem for the model, it is surely not aesthetically appealing. Moreover, on the base of the present analysis, we cannot exclude an instability, similar to that discovered in \cite{BMM}, that turns the black hole spacetime structure in a trapped surface.

The event horizon, if it exists, is given by the largest root of $F(x) = 0$. The right panel in Fig.~\ref{fig1} shows the function $F(x)$. Depending on the value of the mass $M$, there are three possible scenarios. The critical mass separating the three scenarios is
\be
M_{\rm crit} \approx 2.165 \, \frac{M_{\rm Pl}^2}{\sqrt{2} \Lambda} \, .
\ee
For $M < M_{\rm crit}$, $F(x)$ is always positive and there is no horizon. The case $M = M_{\rm crit}$ represents an extremal black hole with a degenerate horizon at the coordinate $x_{\rm crit} \approx 2.063$. For $M > M_{\rm crit}$, $F(x)$ has two roots. The larger root corresponds to an outer event horizon. The smaller root is an inner Cauchy horizon. We can call Lee-Wick black holes the solutions with $M \ge M_{\rm crit}$.

\section{Black hole thermodynamics \label{s-3}}

The temperature of a black hole is $T = \kappa/(2\pi)$, where $\kappa$ is the surface gravity at the event horizon. In the case of a static and spherically symmetric solution with the form~(\ref{eq-ds}), the surface gravity is
\be
\kappa = \frac{1}{2} \frac{dF}{dr} \Big|_{r = r_H} \, ,
\ee
where $r_H$ is the radial coordinate of the event horizon. The temperature of our black holes is
\be
T = - \frac{\Lambda}{4\sqrt{2}\pi} \frac{d}{dx}\left[\ln \frac{f(x)}{x}\right]_{x = x_H} \, .
\ee

\begin{figure*}[t]
%\vspace{0.2cm}
\begin{center}
\vspace{0.8cm}
\boxed{
\includegraphics[width=7.3cm]{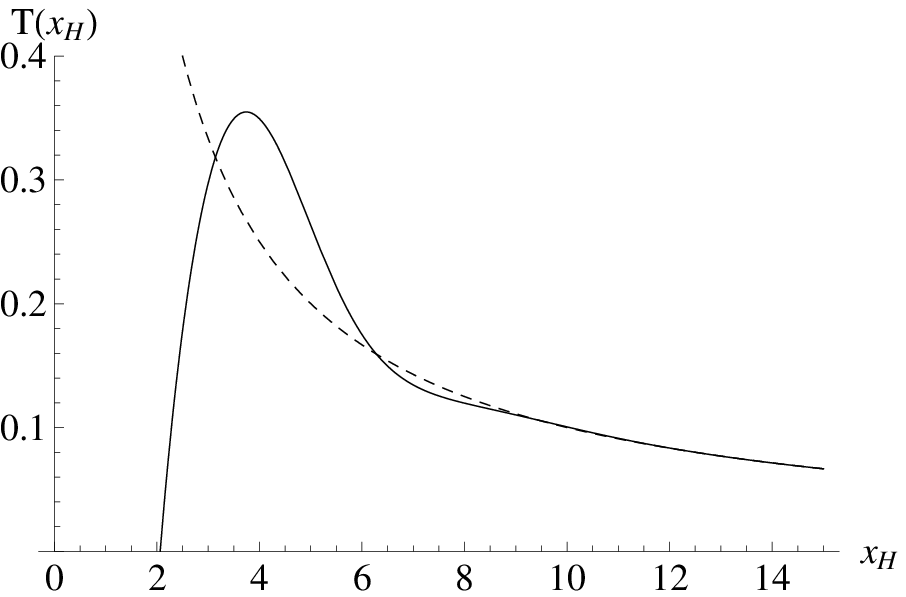}}
\hspace{1.2cm}
\boxed{
\includegraphics[width=8.15cm]{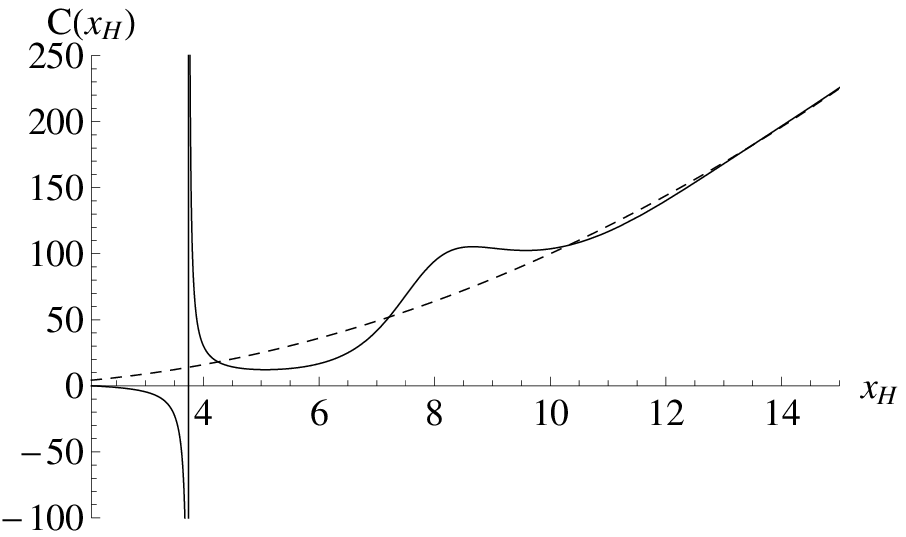}}
\end{center}
\caption{Temperature (left panel) and heat capacity (right panel) of a Lee-Wick black hole (solid line) and of a Schwarzschild black hole (dashed line) as a function of the dimensionless radial coordinate of the even horizon $x_H$. Temperature in units in which $\Lambda/(4 \sqrt{2} \pi) = 1$. Heat capacity in units in which $4\pi/(G_{\rm N} \Lambda^2) = 1$. See the text for more details. \label{fig2}}
\end{figure*}

The left panel in Fig.~\ref{fig2} shows the black hole temperature $T$ as a function of the dimensionless radial coordinate of the even horizon $x_H$. A large $x_H$ corresponds to a black hole with a large mass and we recover the Schwarzschild result. For small $x_H$, we have deviations from the standard picture. The black hole temperature vanishes for $x_{\rm crit} \approx 2.063$, which corresponds to the horizon of the extremal black hole with $M = M_{\rm crit}$. As we will show later, the evaporation process of a black hole leads to a remnant that asymptotically approaches this zero-temperature critical state.

The black hole specific heat is
\be
&& \hspace{-0.6cm}
 C = \frac{dM}{dT} = \frac{dM}{dx_H}\frac{dx_H}{dT} \nonumber\\
&& \hspace{-0.2cm} = - \frac{4\pi}{G_{\rm N} \Lambda^2} 
\frac{d}{dx_H} \left[ \frac{x_H}{f(x_H)} \right] 
\left[ \frac{d^2}{dx_H^2} \ln\frac{f(x_H)}{x_H} \right]^{-1}  \!\! .
\ee
The right panel in Fig.~\ref{fig2} shows the heat capacity as a function of $x_H$. When $x_H \gg 1$, we recover the Schwarzschild limit (dashed line). At $x_H \approx 3.743$, which corresponds to the point of the peak of the temperature in the left panel in Fig.~\ref{fig2}, the heat capacity diverges: this indicates the existence of a phase transition. At smaller $x_H$, the heat capacity approaches zero, and it vanishes at $x_H = x_{\rm crit}$, namely at the extremal black hole state.

\begin{figure*}
\begin{center}
\vspace{0.8cm}
\boxed{
\includegraphics[width=7.8cm]{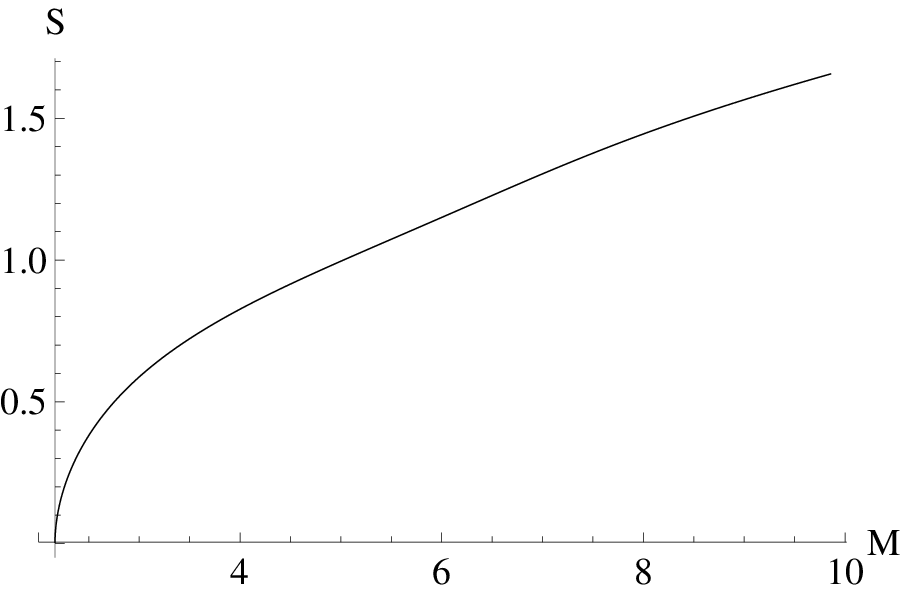}}
\hspace{1.2cm}
\boxed{
\includegraphics[width=7.4cm]{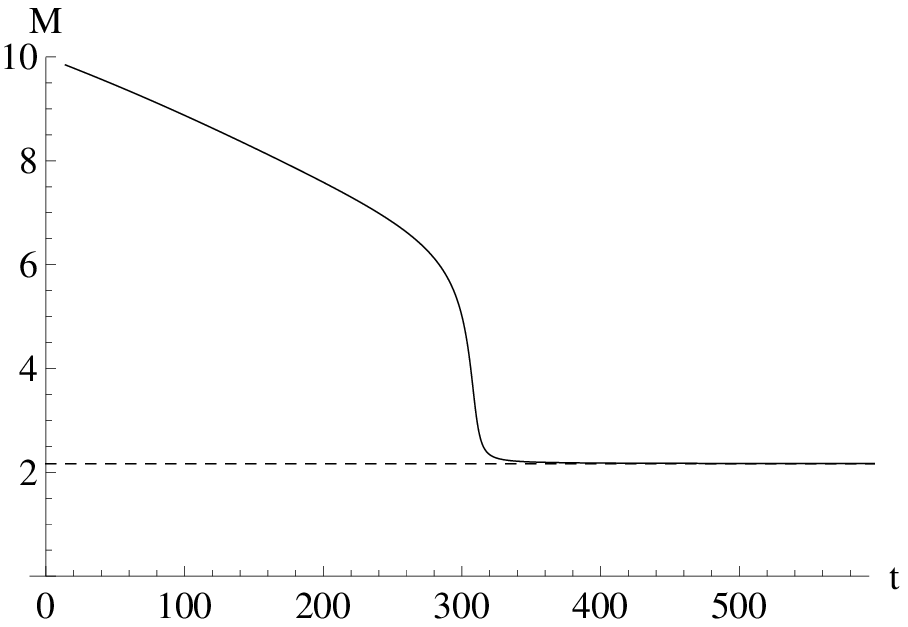}}
\end{center}
\caption{Entropy $S$ in terms of the mass $M$ (left panel) and temporal evolution of $M$ due to the evaporation process (right panel). The remnant approaches the asymptotic state with $M = M_{\rm crit} \approx 2.165 \, M_{\rm Pl}^2 / (\sqrt{2} \Lambda)$ (horizontal dashed line) in an infinite time. Entropy in units in which $4\pi/(G_{\rm N} \Lambda^2) = 1$. See the text for more details.}
\label{fig3}
\end{figure*}

The black hole entropy is
\be
&& S = \int_{r_{\rm crit}}^{r_H} \frac{dM}{T} = 
\int_{x_{\rm crit}}^{x_H} \frac{dM}{dx_H} \frac{dx_H}{T} \\
&& \hspace{0.35cm}
=
\frac{4\pi}{G_{\rm N} \Lambda^2}
\int_{x_{\rm crit}}^{x_H} dx_H \frac{x_H}{f(x_H)} \, .
\ee
The left panel in Fig.~\ref{fig3} shows the entropy as a function of the black hole mass $M$.

The luminosity due to Hawking evaporation can be written as
\be
L = - \frac{dM}{dt} = \sigma A T^4 \, ,
\ee
where $\sigma$ is a parameter whose value depends on the black hole mass and the particle content of the theory, $A = 4 \pi r_H^4$ is the area of the event horizon, and $T$ is the black hole temperature. The black hole evaporation time is thus
\be
\tau_{\rm evap} = \int_i^f dt = \int_i^f \frac{dM}{dx_H} \frac{dx_H}{L} \, .
\ee
The right panel in Fig.~\ref{fig3} shows the evolution in time of the black hole mass $M$ as a consequence of the evaporation process. At $t=0$, the black hole is assumed to have the event horizon at $x_H = 10$. The evaporation process slows down when $M$ is close to the critical value $M_{\rm crit}$, where the black hole temperature vanishes. Therefore, the final product of the evaporation is a remnant that approaches the zero-temperature extremal black hole state in an infinite amount of time
\cite{Nicolini1, Nicolini2, ModestoMoffatNico}.

\section{Conclusions} \label{s-4}

In this letter, we have considered a higher derivative (six order) local modification of Einstein-Hilbert  gravity and we have found solutions of the approximate EOM. Indeed, the approximate EOM can be seen as the EOM of Einstein's gravity coupled to an effective energy-momentum tensor.

We have studied the solution generated by a point-like massive source. There are three qualitatively different scenarios. If the mass of the source assumes a critical value $M_{\rm crit}$, we have an extremal black hole with a degenerate horizon. For $M < M_{\rm crit}$, the solution has no horizon. For $M > M_{\rm crit}$, we have an outer event horizon and an inner Cauchy horizon. However, if $M \gg M_{\rm crit}$ the solution is very close to the Schwarzschild metric.

We have also studied the basic thermodynamics properties of these black holes and their evaporation process. The critical state with $M = M_{\rm crit}$ has a vanishing temperature, while for $M \gg M_{\rm crit}$ the black holes behave like the Schwarzschild one. When $M \rar M_{\rm crit}$, we find some important deviations from the traditional Schwarzschild picture. The evaporation process leads to a remnant, with mass $M \approx M_{\rm crit}$, that approaches the zero-temperature extremal black hole state in an infinite amount of time. 
However, we do not exclude other exact or numerical solutions with the spacetime structure of the gravitational collapse characterized by the formation of a trapped surface instead of a black hole. 
This is a feature of the quadratic gravity \cite{Frolov1,Frolov2} that could be shared by more higher derivative theories too.

%%%%%%%%%%%%%%%%%%%%%%%%%%%%%%%

\begin{acknowledgments}
C.B. was supported by the NSFC (grants 11305038 and U1531117), the Thousand Young Talents Program, and the Alexander von Humboldt Foundation.
\end{acknowledgments}

%%%%%%%%%%%%%%%%%%%%%%%%%%%%%%%

\end{document}